\input epsf
\documentclass[nofootinbib,aps,12pt,preprint,floatfix]{revtex4}

\usepackage{graphicx}
\usepackage{amsmath}
\usepackage{amssymb}
\usepackage{mathrsfs}

\def\Bsmu{B_s \to \mu^+ \mu^-}
\def\Bsmix{B_s - \overline{B}_s}
\def\beq{\begin{equation}}
\def\eeq{\end{equation}}
\def\bey{\begin{eqnarray}}
\def\eey{\end{eqnarray}}
\def\nn{\nonumber}
\def\bdm{\begin{displaymath}}
\def\edm{\end{displaymath}}

\begin{document}
%\preprint{WSU-HEP-0704}
\setcounter{page}{0}
\title{New Physics Contribution to $\Bsmu$ within R-Parity Violating Supersymmetric Models}

\author{Gagik Yeghiyan}

\email{yeghiyag@gvsu.edu}

\affiliation{\vskip 0.3cm Department of Physics, \\
Grand Valley State University, Allendale, Michigan 49401, USA.\vskip 0.05cm}

\abstract{\vskip 0.3cm We re-visit the problem of New Physics (NP) contribution to the branching ratio of the $\Bsmu$ decay in light
of the recent observation of this decay by LHCb. We consider R-parity violating (RPV) supersymmetric models as a primary example -
recently one has reported stringent constraints on the products of the RPV coupling constants that account for the $\Bsmu$ transition
at the tree level.
We argue that despite the LHCb measurement of the $B(\Bsmu)$ is in a remarkable
agreement with the Standard Model (SM) prediction, there is still room for a significant New Physics contribution to the $B(\Bsmu)$,
as the sign of the $\Bsmu$ transition amplitude may be opposite to that
of the Standard Model; alternatively the amplitude may have a large phase.
We conduct our analysis mainly for the case of real RPV couplings.  We find that taking into account the scenario with the sign flip of
the $\Bsmu$ amplitude (as compared to that of the SM) makes the bounds on the RPV coupling products significantly weaker.
Also, we discuss  briefly how our
results are modified if the RPV couplings have large phases. In particular, we examine the dependence of the derived bounds
on the phase of the NP amplitude.} }
\maketitle
\thispagestyle{empty}

%\vspace{0.5cm}

The rare $\Bsmu$ decay is believed to be one of the most powerful tools to test the physics that may occur beyond the Standard Model.
Within the Standard Model this decay is loop-induced and in addition is helicity suppressed. Numerical evaluation gives \cite{1,2,25}
\beq
B(\Bsmu) = \left(3.25 \pm 0.17 \right) \times 10^{-9}  \label{1}
\eeq

In contrast, the $B_s \to \mu^+ \mu^-$ decay rate may be dramatically enhanced within some of the Standard Model
extensions and may exceed the SM prediction by several orders of magnitude. At the same time this decay is characterized by
a pure final leptonic state, which causes the theoretical predictions for it to be very clean. It was therefore used intensively
to constrain the SM extensions, and there was a hope to observe a distinct New Physics signal in this decay mode.

Recently the LHCb collaboration has reported the first evidence for the $\Bsmu$ decay at 3.5$\sigma$ level \cite{3},
\beq
\overline{B}_{exp}(\Bsmu) = \left(3.2^{+1.5}_{-1.2} \right) \times 10^{-9} \label{2}
\eeq
which is in a remarkable agreement with the SM prediction. However, it would not be correct to declare that
there is no New Physics contribution to $\Bsmu$ at all. Some of the popular SM extensions do predict indeed a
negligible NP contribution to the $\Bsmu$ decay rate, due to strong correlations between the $\Bsmix$ mixing and $\Bsmu$
amplitudes \cite{4}. Yet, for other SM extensions the problem of New Physics contribution to $\Bsmu$ in light of the
recent observation of this decay by LHCb is the subject of discussion in the literature \cite{2,5,6,7,8}. In particular,
it has been argued in \cite{5,6,7} that the LHCb result still leaves room for a non-negligible NP contribution, due to the uncertainty in
the experimental value of the $B(\Bsmu)$.

In this paper we examine a source of New Physics contribution to $\Bsmu$ that would be actual even in the idealized limit of zero experimental
and theoretical uncertainties in the $B(\Bsmu)$ and perfect coincidence of the SM prediction with the experimental data. Namely, we consider a possibility
for the $\Bsmu$ transition amplitude to
have a sign opposite to that of the Standard Model or to have a large phase.
The LHCb measurement
of the $\Bsmu$ branching ratio constrains the decay rate, whereas the sign (if it is real) or the phase (if it is complex)
of the transition amplitude remains arbitrary. Thus, it is possible that:
\begin{itemize}
\item{If the amplitude is real (or has a small enough phase so that it may be discarded), one may fit the experimental data for the $B(\Bsmu)$ in particular when
\bey
\nn
&& \hspace{-1.3cm} A^{NP}(\Bsmu) \simeq - 2 A^{SM}(\Bsmu),  \hspace{0.3cm} \text{so that} \\
&& \hspace{-1.3cm}  A(\Bsmu) = A^{SM}(\Bsmu)
+ A^{NP}(\Bsmu) \simeq  - A^{SM}(\Bsmu) \label{3}
\eey}
\item{If instead the NP amplitude has a large phase, one may fit the experimental data for the $B(\Bsmu)$ when
\beq
\left|A^{SM}(\Bsmu) + |A^{NP}(\Bsmu)| e^{i \Phi_{NP}} \right|
\simeq \left|A^{SM}(\Bsmu)\right| \label{4}
\eeq
(if neglecting the SM amplitude phase and using the approximation $B_{exp}(\Bsmu) \approx B^{SM}(\Bsmu)$). Note that Eq.~(\ref{4}) implies
\beq
\hspace{-0.4cm}
- 2 A^{SM}(\Bsmu) < Re\left[A^{NP}(\Bsmu) \right] < 0, \label{5}
\eeq
\beq
\left|Im\left[A^{NP}(\Bsmu) \right]\right| \lesssim  \left| A^{SM}(\Bsmu) \right| \label{6}
\eeq
In particular,
\bey
\nn
\left|Im\left[A^{NP}(\Bsmu) \right]\right| \simeq  \left| A^{SM}(\Bsmu) \right| \hspace{0.3cm} \text{if} \\
Re\left[A^{NP}(\Bsmu) \right] \simeq - A^{SM}(\Bsmu) \label{7}
\eey}
\end{itemize}

One may infer from Eq.'s~(\ref{3})~-~(\ref{7}) that the NP contribution to the $\Bsmu$
transition amplitude is the largest when the amplitude just flips the sign as compared to that of the Standard Model
(rather than getting a large non-trivial phase). So, we will be concentrating
here mainly on the case of a real amplitude, by assuming that the relevant NP parameters are real. We will however discuss at the end
of the paper how our results are modified in presence of large phases of the NP parameters.

Note that the possibility of the $\Bsmu$ amplitude sign flip has already been mentioned in \cite{6} where one considered New Physics models with modified Z-boson couplings to down-type quarks. This possibility has been rejected there, as it is disfavored by the constraints on $Z \to b \bar{b}$. To our best knowledge, there is no reason to disfavor the $\Bsmu$ amplitude sign flip within other SM extensions (in fact it has also been implicitly considered
in \cite{2} within the general analysis
of the NP contribution to $\Bsmu$ in a variety of models, with the amplitude phases varied freely from 0 to $\pi$). In our opinion,
the detailed analysis of the possibility that the $\Bsmu$ amplitude may have a sign opposite to that of the Standard Model (or have a large phase)  may be of great importance,
especially in light of future improvement of the experimental accuracy of measurements
of the $\Bsmu$ branching ratio.

We consider here R-parity violating supersymmetric models with leptonic number violation as a primary example. It has been recently argued \cite{8} that
the remarkable agreement between the LHCb measurement and the SM prediction for the $\Bsmu$ branching ratio implies rigorous constraints on the RPV coupling products
that account for the $\Bsmu$ transition at the tree level. We show that if the $\Bsmu$ transition amplitude
is allowed to have a sign opposite to that of the Standard Model, bounds on the RPV couplings may be by order of magnitude weaker.

The most
general Yukawa superpotential for an explicitly broken R-parity supersymmetric theory may be written as
\beq
W_{\diagup \hspace{-0.23cm} R} = \frac{1}{2} \lambda_{ijk} L_i L_j E^c_k + \lambda^\prime_{ijk} L_i Q_j D^c_k +
\frac{1}{2} \lambda^{\prime \prime}_{ijk} U^c_i D^c_j D^c_k \label{8}
\eeq
Here Q and L denote $SU(2)_L$ doublet quark and lepton superfields, and U, D and E stand for
$SU(2)_L$ singlet up-quark, down-quark and charged lepton superfields. Also, ${i, j, k} =
1, 2, 3$ are generation indices. We shall require baryon number symmetry by setting $\lambda^{\prime \prime}_{ijk}$ to
zero. Also, as mentioned above, we will assume the couplings $\lambda_{ijk}$ and  $\lambda^\prime_{ijk}$ are
real.

Subsequently, the Lagrangian
describing the RPV SUSY contribution to $\Bsmu$ can be written as
\beq
{\cal L}_{\diagup \hspace{-0.23cm} R} = - \Biggl( \lambda^\prime_{i23} \tilde{\nu}_{i_L} \bar{b} P_L s +
\lambda^\prime_{i32} \tilde{\nu}_{i_L} \bar{s} P_L b + \lambda_{i22} \tilde{\nu}_{i_L} \bar{\mu} P_L \mu +
\lambda^\prime_{2k2} \tilde{u}_{k_L} \bar{s} P_L \mu + \lambda^\prime_{2k3} \tilde{u}_{k_L} \bar{b} P_L \mu
+ h.c. \Biggr) \label{9}
\eeq
where $P_{L,R}$ are the helicity projection operators, and we use the notation $P_L = (1 - \gamma_5)/2$.
Note that for the sake of transparency of our analysis, we neglect the transformation of the RPV couplings from the weak
isospin basis to the (s)quark and sneutrino mass basis. (We invoke however to the reader to be cautious when
using the bounds on RPV coupling products derived in this paper. Rigorously speaking, they may be used for the processes involving
down type quark~-~down type quark~-~sneutrino and down type quark~-~up type squark~-~charged lepton transitions only.)

\begin{figure}[t]
\centering
\includegraphics[width=17cm]{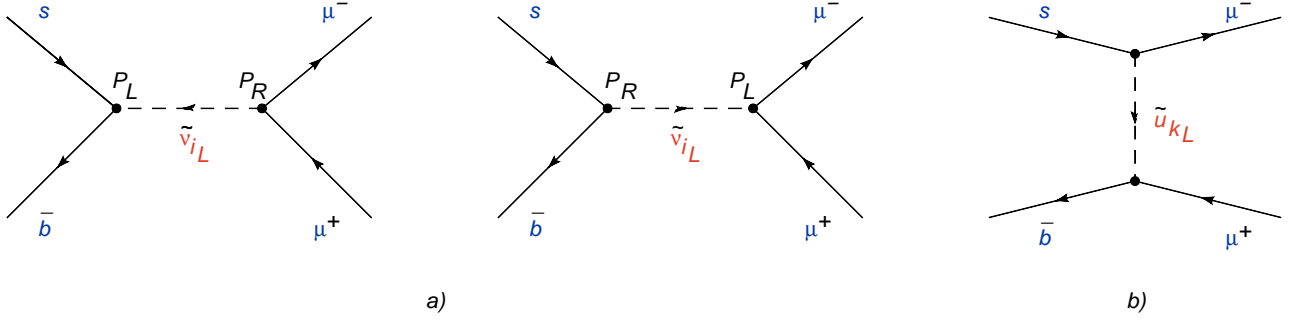}
\caption{Diagrams for the $\Bsmu$ transition within R-parity violating supersymmetric models to the lowest order in perturbation theory, (a) due to exchange
of sneutrinos, (b) due to exchange of up-type squarks. The direction of the sneutrino propagator depends on the helicities of the quark and lepton
states, in other words whether we have $P_L$ or we have $P_R$ operator at an interaction vertex. }
\label{f1}
\end{figure}

Within R-parity violating supersymmetric models, to the lowest order in perturbation theory the $\Bsmu$ transition occurs at the tree level, due to
exchange of sneutrinos or up-type squarks, as depicted in Figure~1. We need also to include the SM contribution to $\Bsmu$: recall that we are interested in destructive interference of the SM and NP amplitudes. Thus, the relevant low-energy $|\Delta B| =1$ effective Hamiltonian
would have the following form:
\beq
H_{eff}^{\Delta B = 1} = H_{eff}^{SM} + H_{eff}^{\tilde{\nu}} + H_{eff}^{\tilde{u}} \label{10}
\eeq
Here \cite{9,10}
\beq
H_{eff}^{SM} = \frac{- 4 G_F}{\sqrt{2}} \frac{\alpha}{2 \pi \sin^2{\theta_W}} \left(V^\star_{t b} V_{t s} \right) \eta_Y Y_0(x_t)
\ \bar{b} \gamma^\nu P_L s \ \bar{\mu} \gamma_\nu P_L \mu  + h.c. \label{11}
\eeq
where \cite{11}
\bdm
Y_0(x_t) = \frac{x_t}{8} \left(\frac{4 - x_t}{1 - x_t} + \frac{3 x_t}{(1 - x_t)^2} \ln{x_t} \right),
\edm
$x_t = m_t^2/M_W^2$,
and $\eta_Y$ is the factor that accounts for the QCD corrections to $Y_0(x_t)$. \\
Two other terms in Eq.~(\ref{10}) are derived by integrating out the sneutrino and squark heavy degrees of freedom. This yields
\beq
H_{eff}^{\tilde{\nu}} = -\left(\frac{\lambda^\star_{i22} \lambda^\prime_{i23}}{m_{\tilde{\nu}_{i_L}}^2} \ \bar{b} P_L s \ \bar{\mu} P_R \mu
\ + \ \frac{\lambda^{\prime^\star}_{i32} \lambda_{i22}}{m_{\tilde{\nu}_{i_L}}^2} \ \bar{b} P_R s \ \bar{\mu} P_L \mu
+ h.c. \right) \label{12}
\eeq
\beq
H_{eff}^{\tilde{u}} = \frac{\lambda^{\prime^\star}_{2k2} \lambda^\prime_{2k3}}{2 m_{\tilde{u}_{k_L}}^2} \ \bar{b} \gamma^\nu P_R s \
\bar{\mu} \gamma_\nu P_L \mu + h.c. \label{13}
\eeq
where $\tilde{\nu}_{i_L}$, $\tilde{u}_{k_L}$ are respectively the lightest sneutrino and the lightest "left" up-type squark
states\footnote{It is assumed that squark mass eigenstates do not differ significantly from the "left" and "right" squark states.
This is known to be the case for most SUSY scenarios with the squark masses much greater than 100~GeV.}. For  (nearly) degenerate sneutrino and/or squark masses, one should replace
in Eq.~(\ref{12}) and/or Eq.~(\ref{13}) the lightest sparticle masses by universal sneutrino and/or squark masses,
$m_{\tilde{\nu}_{i_L}} \to m_{\tilde{\nu}_L}$, $m_{\tilde{u}_{k_L}} \to m_{\tilde{u}_L}$, as well as sum over indices $i$ and/or $k$.

Also, for the sake of clarity of our analysis, we will follow ref.~\cite{4} and assume
\beq
\lambda^\prime_{i23} = \lambda^\prime_{i32} \label{14}
\eeq
in our further calculations.

Using Eq.'s~(\ref{10})~-~(\ref{13}) as well as the simplifying assumption (\ref{14}), one may present the $\Bsmu$ transition amplitude
in the following form:
\beq
A(\Bsmu) = A^{SM}(\Bsmu) + A^{\tilde{\nu}}(\Bsmu) + A^{\tilde{u}}(\Bsmu)   \label{15}
\eeq
where
\beq
\hspace{-0.7cm} A^{SM}(\Bsmu) = - \langle \mu^+ \mu ^- | H_{eff}^{SM} | B_s \rangle
= \frac{- i G_F}{\sqrt{2}} \frac{\alpha f_{B_s} m_\mu}{\pi \sin^2{\theta_W}} \left(V^\star_{t b} V_{t s} \right) \eta_Y Y_0(x_t)
\ \bar{u}(p_-) \gamma_5 {\it v}(p_+)  \label{16}
\eeq
\beq
A^{\tilde{\nu}}(\Bsmu) = - \langle \mu^+ \mu ^- | H_{eff}^{\tilde{\nu}} | B_s \rangle =
\frac{-i \lambda^\star_{i22} \lambda^\prime_{i23} f_{B_s} M_{B_s}^2 }{2 m_{\tilde{\nu}_{i_L}}^2 m_b} \
\bar{u}(p_-) \gamma_5 {\it v}(p_+)
 \label{17}
\eeq
\beq
A^{\tilde{u}}(\Bsmu) = - \langle \mu^+ \mu ^- | H_{eff}^{\tilde{u}} | B_s \rangle =
\frac{-i \lambda^{\prime^\star}_{2k2} \lambda^\prime_{2k3} f_{B_s} m_\mu}{4 m_{\tilde{u}_{k_L}}^2} \
\bar{u}(p_-) \gamma_5 {\it v}(p_+) \label{18}
\eeq
where $u(p_-)$ and ${\it v}(p_+)$ are the bi-spinor wave functions of the leptonic states. (Subsequently,
$p_+$ and $p_-$ are the momenta of $\mu^+$ and $\mu^-$.)
In deriving (\ref{16})~-~(\ref{18}) we used the following parametrization of the hadronic matrix elements:
\begin{eqnarray*}
&& \langle 0 | \bar{b} \gamma^\nu P_L s | B_s \rangle = \frac{i f_{B_s}}{2} p_B^\nu \hspace{1.2cm}
\langle 0 | \bar{b} \gamma^\nu P_R s | B_s \rangle = \frac{- i f_{B_s}}{2} p_B^\nu  \\
&& \langle 0 | \bar{b} P_R s | B_s \rangle = \frac{i f_{B_s}}{2} \left(\frac{M_{B_s}^2}{m_b}\right) \hspace{0.5cm}
\langle 0 | \bar{b} P_L s | B_s \rangle = \frac{- i f_{B_s}}{2} \left(\frac{M_{B_s}^2}{m_b} \right)
\end{eqnarray*}
where $f_{B_s}$ is the $B_s$ meson decay constant, and $p_B$ is the $B_s$ 4-momentum.

We want to stress that all three parts of the amplitude have the same structure. They all contain the same
pseudoscalar bi-spinor bilinear form multiplied by some factor. In what follows, both $A^{\tilde{\nu}}(\Bsmu)$
and $A^{\tilde{u}}(\Bsmu)$ may interfere with the SM amplitude. In other words, the $\Bsmu$ amplitude
may have a sign opposite to that of the SM both due to the contribution of the sneutrino-mediated diagrams,
and due to the contribution of the squark-mediated diagram\footnote{If we give up the simplifying assumption (\ref{14}),
$A^{\tilde{\nu}}(\Bsmu)$ will also contain a term with a scalar bi-spinor bilinear form. This term however won't
interfere with the other terms of the transition amplitude, so it does not play any essential role in our
analysis.}.

Calculation of the decay branching ratio using (\ref{15})~-~(\ref{18}) is straightforward and yields
\bey
\nn
B(\Bsmu) = \frac{\tau_{B_s} M_{B_s} f_{B_s}^2}{8 \pi} \sqrt{1 - \frac{4 m_\mu^2}{M_{B_s}^2}} \ \Biggl|
\frac{G_F}{\sqrt{2}} \frac{\alpha \ m_\mu}{\pi \sin^2{\theta_W}} \left(V^\star_{t b} V_{t s} \right) \eta_Y Y_0(x_t) \\
+ \ \frac{\lambda^\star_{i22} \lambda^\prime_{i23} \ M_{B_s}^2 }{2 m_{\tilde{\nu}_{i_L}}^2 m_b} \
+ \ \frac{\lambda^{\prime^\star}_{2k2} \lambda^\prime_{2k3} \ m_\mu}{4 m_{\tilde{u}_{k_L}}^2} \Biggr|^2 \label{19}
\eey
where $\tau_{B_s}$ is the average lifetime of the $B_s$ meson.

We use during the numerical analysis $\tau_{B_s} = 1.509$~ps, \cite{12}, $f_{B_s} = 0.225$~GeV \cite{13},
$V^\star_{tb} V_{ts} = 0.0405$ \cite{19} (as mentioned above, we neglect the small phase of this CKM product),
$M_W = 80.4$~GeV, $\sin^2{\theta_W} = 0.231$, $G_F = 1.166 \times 10^{-5}~GeV^{-2}$, $\alpha = \alpha(M_Z) = 1/128$,
$m_\mu = 0.106$~GeV, $M_{B_s} = 5.3667$~GeV, $m_b = \overline{m}_b(m_b) = 4.18$~GeV \cite{14}.
For the top quark mass we use $m_t^{pole} = 173.2$~GeV \cite{15}, which yields for the $\overline{MS}$,QCD renormalized mass
$\overline{m}_t(m_t) = 163.2$~GeV \cite{1}; $\eta_Y = 1.012$ for $x_t = \overline{m}_t^2(m_t)/M_W^2$ \cite{1,16,17,18}.

We neglect the uncertainties in the values of the input parameters specified above. Those are known to alter the predictions
for the $B(\Bsmu)$ by about 10\% \cite{1,2}. This uncertainty in the $B(\Bsmu)$  is much less than the one in the experimental value of the
branching ratio and the one in our results due to destructive interference of different NP amplitudes (see the discussion at the
end of the paper).

We choose $m_{\tilde{\nu}_{i_L}} \gtrsim 100$~GeV and $m_{\tilde{u}_{k_L}} \gtrsim 500$~GeV. The squark masses below 500~GeV
are highly disfavored by the LHC data (see \cite{14,26} and references therein). To our best knowledge, however, no such strong
constraints on sneutrino masses has been derived so far \cite{14}.

Also, following the common approach, we will assume only one non-vanishing RPV coupling product at a time, or alternatively
only one of the NP amplitudes in (\ref{15}) to be non-vanishing at a time.

We consider first an idealized scenario with zero uncertainties in the experimental and theoretical values of the $\Bsmu$ branching
ratio and perfect coincidence of the Standard Model prediction with the experimental data. In such a scenario (if assuming
real RPV couplings), non-vanishing New Physics contribution to $\Bsmu$ may occur if only the transition amplitude has a sign
opposite to that of the Standard Model. Following the approach of one non-vanishing coupling product at a time, we choose first
$\lambda^{\prime^\star}_{2k2} \lambda^\prime_{2k3} = 0$ or equivalently $A^{\tilde{u}}(\Bsmu) = 0$. Then the
transition amplitude flips the sign if
\bdm
A^{\tilde{\nu}}(\Bsmu) = - 2 A^{SM} (\Bsmu)
\edm
Using Eq.'s~(\ref{16}) and (\ref{17}) and the values of the input parameters specified above, one finds that this occurs when
\beq
- \lambda^\star_{i22} \lambda^\prime_{i23} = 2.12 \times 10^{-6} \left(\frac{m_{\tilde{\nu}_{i_L}}}{100~GeV} \right)^2 \label{20}
\eeq
This value of $\lambda^\star_{i22} \lambda^\prime_{i23}$ is several times greater in magnitude than the bound quoted in \cite{8}
(as no amplitude sign flip or large phase has been considered in \cite{8}). Nevertheless, Eq.~(\ref{20}) implies rigorous constraints
on this coupling product or alternatively on the sneutrino masses. Indeed, Eq.~(\ref{20}) implies
$(- \lambda^\star_{i22} \lambda^\prime_{i23}) \sim 10^{-6}$ for the lightest sneutrino mass  $\sim 100$~GeV. Alternatively, if one desires
for this coupling product to be of the same order as the SM weak coupling squared ($g^2 \sim 0.5$), the lightest sneutrino should have a
mass $\sim 50$~TeV. This is a manifestation of the so-called flavor problem \cite{20,21}: to assure that tree level flavor changing neutral currents
beyond the SM do not conflict with the experimental data, either the relevant couplings should be unnaturally small or the New Physics
mass scale should be enormously large. Solving the flavor problem goes beyond the scope of the present paper. Instead we will simply assume further that
$\lambda^\star_{i22} \lambda^\prime_{i23} = 0$, or $A^{\tilde{\nu}}(\Bsmu)$ vanishes, and
we will be concentrating on the contribution of the squark-mediated diagram only (Fig.~\ref{f1}~(b)). As mentioned above, possible effects of
interference of different NP amplitudes will be discussed at the end of the paper.

If assuming $A^{\tilde{\nu}}(\Bsmu) = 0$, the transition amplitude flips the sign when
\bdm
A^{\tilde{u}}(\Bsmu) = - 2 A^{SM} (\Bsmu)
\edm
Using Eq.'s~(\ref{16}) and (\ref{18}) and the values of the input parameters specified above, one finds that this occurs when
\beq
- \lambda^{\prime^\star}_{2k2} \lambda^\prime_{2k3} = 6.88 \times 10^{-3} \left(\frac{m_{\tilde{u}_{k_L}}}{500~GeV} \right)^2 \label{21}
\eeq
Eq.~(\ref{21}) implies rather weak constraints on the couplings $\lambda^{\prime}_{2k2}$ and $\lambda^\prime_{2k3}$. If
assuming no hierarchy in the values of $\lambda^\prime_{2k2}$ and $\lambda^\prime_{2k3}$, one gets
$|\lambda^\prime_{2k2}| \sim 0.085$ and $|\lambda^\prime_{2k3}| \sim 0.085$ for $m_{\tilde{u}_{k_L}} \sim 500$~GeV. Thus,
moderately small values of $\lambda^\prime_{2k2}$
and $\lambda^\prime_{2k3}$ are still allowed for $m_{\tilde{u}_{k_L}} \sim 500$~GeV. Furthermore, choosing
the lightest "left" up-type squark mass to be heavier (say 1~TeV of few~TeV) would yield larger values for $\lambda^{\prime}_{2k2}$ and $\lambda^\prime_{2k3}$ (and for their product) to be allowed.

At first glance this result is not
surprising, as the contribution of the diagram with a squark exchange in Fig.~\ref{f1}~(b) is helicity suppressed (like the SM contribution), as can
be seen e.g. from Eq.~(\ref{18}).
We want to stress however
that this  is a rather non-trivial result, in a sense that one should consider the possibility of  the $\Bsmu$ amplitude sign flip to
derive it. If instead one assumes that the sign of the transition amplitude is the same as within the SM, so that the NP contribution
is solely due to the uncertainty in the experimental value of the $B(\Bsmu)$, the constraints on the coupling product
$\lambda^{\prime^\star}_{2k2} \lambda^\prime_{2k3}$ are significantly stronger. To illustrate this, we will consider a realistic scenario now: we will
demand that our predictions for the $\Bsmu$ branching ratio fall in the experimentally allowed interval.

In order to do this, one should take into account that the experimentally measured branching ratio of the $\Bsmu$ decay is the time integrated
branching ratio (usually denoted $\overline{B}(\Bsmu)$ like in Eq.~(\ref{2}) above). It is related to the "theoretical" branching ratio as \cite{2, 27, 22, 23}:
\beq
B(\Bsmu) = \left(\frac{1 - y_s^2}{1 + A_{\Delta \Gamma}^{\mu \mu} y_s} \right) \overline{B}(\Bsmu) \label{22}
\eeq
Here \cite{24}
\beq
y_s = \frac{\Delta \Gamma_s}{2 \Gamma_s} = 0.088 \pm 0.014 \label{23}
\eeq
where $\Delta \Gamma_s$ is the width difference in the $\Bsmix$ mixing, and $\Gamma_s$ is the average width of the $B_s$ meson. The expression for
$A_{\Delta \Gamma}^{\mu \mu}$ in terms of Wilson coefficients of the low-energy effective operators may be found in \cite{2}. For the considered case of real
NP couplings and under simplifying assumption (\ref{14}), one can show after doing some algebra that $A_{\Delta \Gamma}^{\mu \mu} = 1$. Thus, the experimentally allowed (1$\sigma$) interval for the $\overline{B}(\Bsmu)$ (given by Eq.~(\ref{2})) is converted to the following allowed interval for the theoretical branching ratio:
\beq
B(\Bsmu) = (1 - y_s) \overline{B}(\Bsmu) = \left(2.9^{+1.4}_{-1.2}\right) \times 10^{-9}  \label{24}
\eeq

Eq.~(\ref{24}) (combined with Eq.~(\ref{19}) in the limit when only the squark mediated diagram in Fig.~\ref{f1}~(b) gives a non-vanishing NP contribution) yields the following constraints on
the coupling product $\lambda^{\prime^\star}_{2k2} \lambda^\prime_{2k3}$:
\beq
-4.9 \times 10^{-4} \left(\frac{m_{\tilde{u}_{k_L}}}{500~GeV} \right)^2 \leq - \lambda^{\prime^\star}_{2k2} \lambda^\prime_{2k3}
\leq 9.6 \times 10^{-4} \left(\frac{m_{\tilde{u}_{k_L}}}{500~GeV} \right)^2 \label{25}
\eeq
and
\beq
5.92 \times 10^{-3} \left(\frac{m_{\tilde{u}_{k_L}}}{500~GeV} \right)^2 \leq - \lambda^{\prime^\star}_{2k2} \lambda^\prime_{2k3}
\leq 7.37 \times 10^{-3} \left(\frac{m_{\tilde{u}_{k_L}}}{500~GeV} \right)^2 \label{26}
\eeq
The first interval (given by (\ref{25})) is derived, when the $\Bsmu$ transition amplitude has the same sign as that of the Standard Model. The New Physics
contribution is due to the uncertainty in the experimental value of the $\Bsmu$ branching ratio. This interval for $\lambda^{\prime^\star}_{2k2} \lambda^\prime_{2k3}$
is in a reasonable agreement with that quoted in ref.~\cite{8}. The second interval (given by (\ref{26})) is derived when the transition
amplitude has a sign opposite to that of the Standard Model. In that case the allowed values of
$-\lambda^{\prime^\star}_{2k2} \lambda^\prime_{2k3}$ are greater by an order of magnitude. As discussed above, this implies weaker constraints on
the allowed region of the NP parameter space.

Notice also that for the $\Bsmu$ amplitude to flip the sign, the coupling product $\lambda^{\prime^\star}_{2k2} \lambda^\prime_{2k3}$
must be negative (as it follows from Eq.~(\ref{26})). Contrary to this, within the other interval (given by (\ref{25})), the sign of
$\lambda^{\prime^\star}_{2k2} \lambda^\prime_{2k3}$ is arbitrary.

We used the 1$\sigma$ experimental interval to derive the constraints on
$\lambda^{\prime^\star}_{2k2} \lambda^\prime_{2k3}$ given by (\ref{25}) and (\ref{26}). A more conservative approach would imply using the 95\%~C.L.
interval, $\overline{B}(\Bsmu) = [1.1 \div 6.4] \times 10^{-9}$ \cite{3}. One would observe the same effect in that case as well, although less pronounced
and harder to analyze. While using the 95\%~C.L. interval (instead of the 1$\sigma$ one) would affect the sign-flip interval (given by (\ref{26})) by about 10\% only, the same-sign interval would be significantly more wide-spread than (\ref{25}). We leave for a reader to verify that if using the 95\%~C.L.
interval, the maximum value of $- \lambda^{\prime^\star}_{2k2} \lambda^\prime_{2k3}$ in the sign-flip interval would be about five times greater than
the maximum value of $|\lambda^{\prime^\star}_{2k2} \lambda^\prime_{2k3}|$ in the same-sign interval, or constraints on this coupling product would still be significantly weaker when taking into account the possibility of the $\Bsmu$ amplitude sign flip.

In principle, one may conduct a similar analysis for the contribution of the sneutrino-mediated diagrams in Fig.~\ref{1}~(a) and subsequently for the other coupling product, $\lambda^\star_{i22} \lambda^\prime_{i23}$. Assuming now
that the squark-mediated diagram in Fig.~\ref{1}~(b) has a vanishing contribution to $\Bsmu$, one will get in this case two different intervals
for $\lambda^\star_{i22} \lambda^\prime_{i23}$
(that originate in the same way as (\ref{25}) and (\ref{26}) for $\lambda^{\prime^\star}_{2k2} \lambda^\prime_{2k3}$). We leave this for a reader
as another exercise to do.

In a realistic scenario neither of the diagrams in Fig.~\ref{f1} may have a vanishing contribution to $\Bsmu$. In addition, one should also
take into account the impact of the R-conserving sector of the theory on the $\Bsmu$ transition amplitude as well \cite{5}.
Thus, in a realistic scenario one has different
sources of a NP contribution to $\Bsmu$ that may in general interfere both constructively and destructively \cite{28}.

If different NP amplitudes interfere constructively, the coupling product $\lambda^{\prime^\star}_{2k2} \lambda^\prime_{2k3}$ may also
acquire the values between the two intervals given by (\ref{25}) and (\ref{26}). (That is to say, the $\Bsmu$ amplitude sign flip
may be only in
part due to the contribution of the squark-mediated diagram in Fig.~\ref{f1}~(b), it may also be in part due to other New Physics effects.)
In other words, one should replace (\ref{25}) and (\ref{26}) by
\beq
-4.9 \times 10^{-4} \left(\frac{m_{\tilde{u}_{k_L}}}{500~GeV} \right)^2 \leq  - \lambda^{\prime^\star}_{2k2} \lambda^\prime_{2k3}
\leq 7.37 \times 10^{-3} \left(\frac{m_{\tilde{u}_{k_L}}}{500~GeV} \right)^2 \label{27}
\eeq

Of course, the NP amplitudes may also interfere destructively. In that case the bounds on $\lambda^{\prime^\star}_{2k2} \lambda^\prime_{2k3}$
given by Eq.~(\ref{27}) may somehow be distorted (they may become weaker). Yet, if there is no fine-tuning or exact cancellation of the contributions
of different NP amplitudes, it is very unlikely that this distortion alter the bounds on $\lambda^{\prime^\star}_{2k2} \lambda^\prime_{2k3}$,
say, by an order of magnitude. Thus, one may always use (\ref{27}) to get an insight into how large (in order of magnitude) the
coupling product $\lambda^{\prime^\star}_{2k2} \lambda^\prime_{2k3}$ is still allowed to be.

\begin{figure}[t]
\includegraphics[width=16cm]{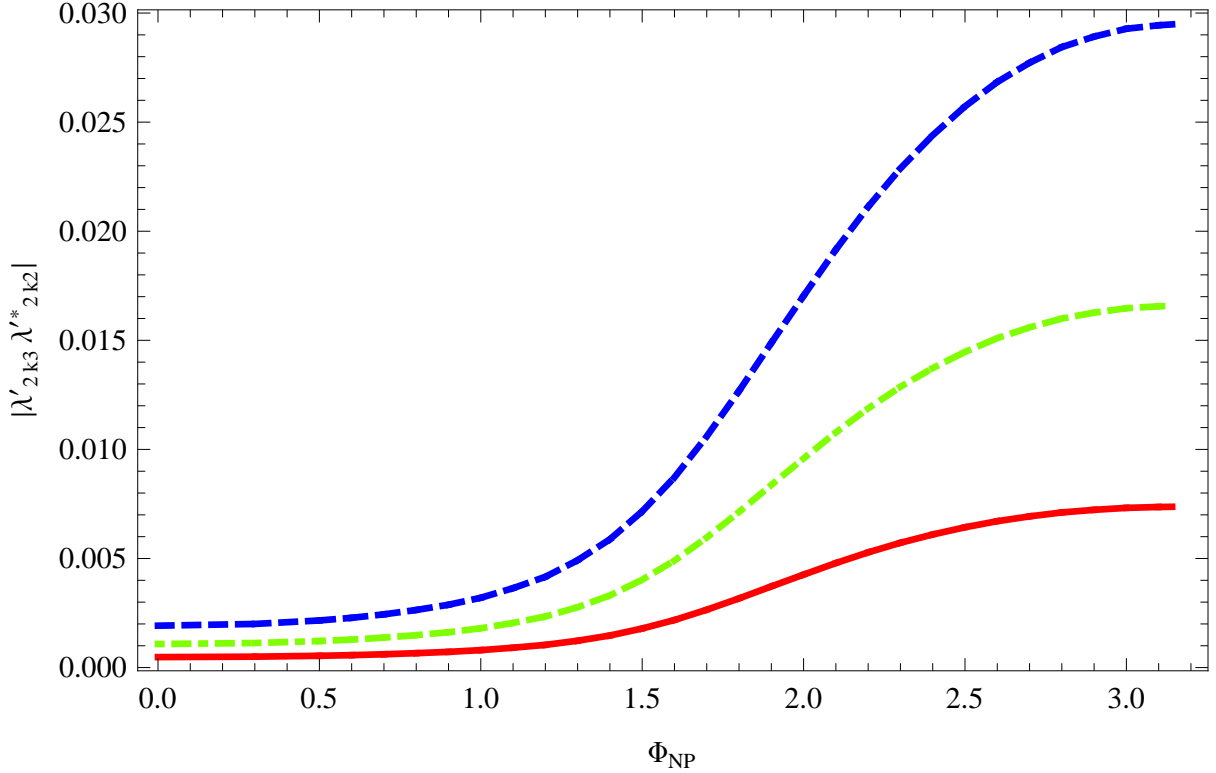}
\caption{Upper bound on $|\lambda^{\prime \star}_{2k2} \lambda^\prime_{2k3}|$ as a function of the NP amplitude phase $\Phi_{NP}$ for
$m_{\tilde{u}_{k_L}} = 500$~GeV (solid red line), $m_{\tilde{u}_{k_L}} = 750$~GeV (dashed-dotted green line),
$m_{\tilde{u}_{k_L}} = 1$~TeV (dashed blue line).}
\label{f2}
\end{figure}

So far we were assuming that the R-parity violating couplings are real (or have small enough phases so that they may be discarded). Yet, our analysis may be extended also to the case when
these couplings have large phases. Demanding again that our predictions fall into the experimentally allowed
interval (and assuming again $A^{\tilde{u}}(\Bsmu)$ to be the only non-vanishing NP amplitude), one may derive an upper bound on the absolute value of the
coupling product $\lambda^{\prime \star}_{2k2} \lambda^\prime_{2k3}$ as a function of the NP amplitude phase $\Phi_{NP} = \arg\left(\lambda^{\prime \star}_{2k2} \lambda^\prime_{2k3}\right)$. One must however be cautious what the allowed interval is now, as the observable $A_{\Delta \Gamma}^{\mu \mu}$ is not equal to unity anymore. Thus, Eq.~(\ref{24}), which we were using in the case of real NP couplings, is not valid here. One shall demand instead
\beq
\left(\frac{1 + A_{\Delta \Gamma}^{\mu \mu} y_s}{1 - y_s^2} \right) B(\Bsmu) = \overline{B}(\Bsmu) = \left(3.2^{+1.5}_{-1.2} \right) \times 10^{-9} \label{29}
\eeq
where in the limit of vanishing contribution of scalar operators, $A_{\Delta \Gamma}^{\mu \mu}= \cos(2 \varphi_P - \phi_s^{NP})$\cite{2}. Here $\phi_s^{NP}$ is the NP piece of the $\Bsmix$ mixing phase, and within the considered scenario $\varphi_P$ is the phase of the total $(SM + NP)$ amplitude. There is a rather weak correlation between the NP contribution
to $\Bsmix$ coming from the R-parity violating sector and that to $\Bsmu$ \cite{4}. Moreover, this correlation is negligible, if analyzing the contribution of the squark-mediated diagram in Fig.~\ref{f1}~(b) only (or analyzing the constraints on $\lambda^{\prime \star}_{2k2} \lambda^\prime_{2k3}$). Also, the recent measurements of $\phi_s$
at LHCb \cite{24}, combined with the knowledge of the SM piece of $\phi_s$, allow us to infer that $\phi_s^{NP} \lesssim 0.15$~radians, so this phase is too small to affect $A_{\Delta \Gamma}^{\mu \mu}$ significantly. We will discard $\phi_s^{NP}$ in our calculations, thus using $A_{\Delta \Gamma}^{\mu \mu} \approx \cos{2 \varphi_P}$. Note that $\varphi_P$ does not acquire a unique value as the NP amplitude phase $\Phi_{NP}$ is fixed. $\varphi_P$ depends both on $\Phi_{NP} = \arg\left(\lambda^{\prime \star}_{2k2} \lambda^\prime_{2k3}\right)$ (which is the only genuine free parameter in our analysis), and on $|\lambda^{\prime \star}_{2k2} \lambda^\prime_{2k3}|/m_{\tilde{u}_{k_L}}^2$ (or on the relative weight of the NP amplitude compared to the SM one).

The derived bound on $|\lambda^{\prime \star}_{2k2} \lambda^\prime_{2k3}|$ as a function of the NP amplitude phase $\Phi_{NP}$ is presented in Fig.~\ref{f2}. As one can see from Fig.~\ref{f2}, the bound on $|\lambda^{\prime \star}_{2k2} \lambda^\prime_{2k3}|$ becomes weaker as $\Phi_{NP}$
gets larger, and it is the weakest when $\Phi_{NP} \to \pi$. As mentioned above, this result could also be inferred from the analysis of Eq.'s~(\ref{3})~-~(\ref{7}). Thus, the most general bound on $|\lambda^{\prime \star}_{2k2} \lambda^\prime_{2k3}|$ (both in the case when this product
is real and in the case this product has a phase) would be
\beq
\left|\lambda^{\prime^\star}_{2k2} \lambda^\prime_{2k3} \right|
\lesssim 7.37 \times 10^{-3} \left(\frac{m_{\tilde{u}_{k_L}}}{500~GeV} \right)^2 \label{28}
\eeq

In conclusion, we have re-visited the problem of New Physics contribution to the $\Bsmu$ decay in light of the recent experimental measurement
of this decay branching ratio by the LHCb collaboration. We have examined R-parity violating supersymmetric models
as a primary example, and argued that there is still room for a significant NP contribution, as the transition amplitude still
may have a sign opposite to that of the Standard Model or alternatively may get a large phase. We have found that if taking into account the
effect of the $\Bsmu$ amplitude possible sign flip as compared to that of the SM (or possible large phase), the
bounds imposed on the RPV coupling products that account for the $\Bsmu$ transition may be weaker by an order of magnitude
than if the effect of the amplitude sign flip (or possible large phase) is disregarded. We emphasize that a similar effect may be
observed also within other SM extensions. So considering within other New Physics models the possibility for the
$\Bsmu$ transition amplitude to have a sign
opposite to that of the SM or have a large non-trivial phase is strongly encouraged. \\

The author is grateful to Alexey A. Petrov and Javier  Virto for stimulating discussions and valuable suggestions and comments.

\end{document}